# Numerical studies of long-term wettability alteration effects in $CO_2$ storage applications

D. Landa-Marbán[*,1], K. Kumar[2], S.E. Gasda[1,2], T.H. Sandve[1], A.M. Kassa[1,2]
[1] NORCE Norwegian Research Centre AS, Bergen, Norway
[2] University of Bergen, Bergen, Norway

*Corresponding author e-mail: dmar@norceresearch.no

**Summary**

The wettability of the rock surface in porous media has an effect on the constitutive saturation functions that govern capillary pressure and relative permeability. The term wettability alteration refers to the change of this property over time by processes such as $CO_2$ interactions with the rock. In this work, we perform numerical simulations considering a two-phase two-component flow model including time-dependent wettability alteration in a two-dimensional aquifer-caprock system using the open porous media framework. Particularly, we study the spatial distribution over time of injected $CO_2$ into the aquifer neglecting and including wettability alteration effects.

The numerical simulations show that wettability alteration on the caprock results in a loss of containment; however, the $CO_2$ front into the caprock advances very slow since the unexposed caprock along the vertical migration path also needs to be changed by the slow wettability alteration process. The simulations also show that wettability alteration on the aquifer results in an enhancement of storage efficiency; this since the $CO_2$ front migrates more slowly and the capillary entry pressure decreases after wettability alteration.

**Introduction**

Geological carbon sequestration in deep saline aquifers is considered one of the most promising scalable solutions to reduce anthropogenic $CO_2$ emissions. Examples are Utsira aquifer storage in the North Sea and the geological storage project in Alberta. A critical issue in geological storage of $CO_2$ is to ensure that it will remain stored on the long term in the formation where it is injected. In the time scale of storage of possibly thousands of years, two factors will determine the efficacy of the safe storage. One is the amount of $CO_2$ dissolved and precipitated on the rock surface and the second and the more dominant factor is the buoyant $CO_2$ phase rising above the other fluids but remaining trapped beneath a low-permeable caprock. Consequently, an important factor in the successful retention of geological carbon storage is the capillary breakthrough and the hydraulic properties of the $CO_2$ phase.

The capillary sealing efficiency of the caprock is a function of interfacial tension between the brine and the $CO_2$ and the contact angle with the rock surface. On the macroscale, this is manifested in the form of hydraulic functions of the fluids: capillary pressure and relative permeability of the $CO_2$ and brine in the storage reservoir and the caprock. These functions are in turn determined by pore-scale properties including wettability that controls the contact angle. The term wettability refers to the preference of one fluid over the other for the rock surface and is a pore-scale phenomenon that has direct influence on macroscale displacement processes. The standard assumption is that wettability is a static property of the fluids involved. However, experimental evidence suggests that exposure of the rock surface by the fluid leads to wettability alteration (WA) via a change in the contact angle (Wang et al. 2013; Seyyedi et al. 2015; Fatah et al. 2021; Gholami et al. 2021). This change in the contact angle may lead to change in the residual saturation of the fluids. Recently, Kassa et al. (2020) studied the dynamic capillary pressure that considers the WA due to exposure of the rock to $CO_2$ leading to irreversible change in the wettability of the surface. This has been extended to derive dynamic relative permeability functions in Kassa et al. (2021). The basic idea is to define a new time-dependent variable, called upscaled exposure time, and use this as a parameter in the hydraulic functions. The resulting dynamic capillary functions and relative permeability functions are extended forms of standard constitutive models either by interpolation between two end-state capillary functions or by incorporating WA dynamics directly into the parameters of the existing relative permeability model. The standard forms of capillary functions and relative permeability ensure that they are easy to implement within standard flow simulation toolboxes.

In this work, we further the study of WA through implementation of dynamic models into field-scale simulations. This is an extension of the work in Kassa et al. (submitted) where we studied the impact on wettability alteration in simplified one-dimensional domains. Our approach here is to consider two immiscible fluids $CO_2$ and brine that are mutually soluble fluids. The soluble $CO_2$ can induce chemical alteration of the native rock causing the transformation from a strongly wet condition to an intermediate wet condition. As stated above, we use the $CO_2$ exposure as a variable determining the corresponding hydraulic functions. Since there is a different history of exposure of $CO_2$ to the rock surface in space and time, the extent of WA and the consequent change in the hydraulic functions will vary in space and time. This spatial and temporal complexity in capillary pressure and relative permeability functions can impact field-scale flow behavior in a complex manner. This field-scale study here takes place in two dimension (see Figure 1).

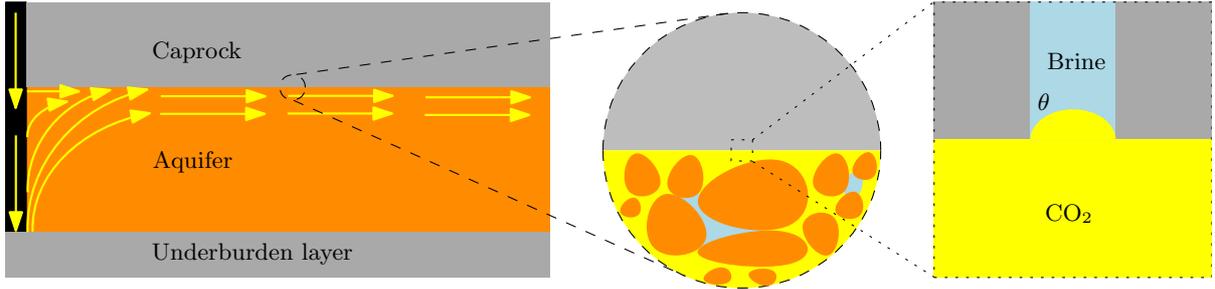

Figure 1: Schematic representation of the 2D aquifer-caprock system including WA effects.

**Methodology**

A comprehensive presentation of the TPTC flow methodology and model including time-dependant WA effects can be found in Kassa et al. (submitted). Here we shortly describe the model. We consider a constant-temperature reservoir initially saturated with brine (wetting phase, $S_w$) where $CO_2$ is injected to be stored (non-wetting phase, $S_n = 1 - S_w$). The injected $CO_2$ dissolves into the brine ($CO_2$ mass fraction in the wetting phase, $X_w^{CO_2} = 1 - X_n^{CO_2}$) which leads to chemical alteration of the rock matrix. This wettability alteration of the rock results in changes on the saturation functions (capillary pressure, $P_c = P_n - P_w$; relative permeabilities, $k_{rw}$ and $k_{rn}$). We model this by including a new macroscale variable (upscaled exposure time, $\bar{\chi}$) and parameter (pore-scale parameter, $C$) into these saturation functions.

Let $t_{ch}$ be a pre-specified characteristic time for WA process. The upscaled exposure time is given as:

$$\bar{\chi} := \frac{1}{t_{ch}} \int_0^t X_w^{CO_2} \, d\tau. \qquad (1)$$

Let $b_1$ and $b_2$ be fitting parameters for capillary pressure changes. The capillary pressure is given as:

$$P_c = \frac{\bar{\chi} S_{ew}}{b_1 C^{b_2} + \bar{\chi} S_{ew}} \left(P_c^f - P_c^i\right) + P_c^i, \qquad (2)$$

where the effective wetting saturation is calculated using the residual phase saturations ($S_{ew} = [S_w - S_{rw}]/[1 - S_{rw} - S_{rn}]$) and the initial and final static capillary pressures are modeled by Brooks-Corey relationships ($P_c^i = c^i S_{ew}^{-1/\lambda}$ and $P_c^f = c^f S_{ew}^{-1/\lambda}$, where $\lambda$ is a fitting parameter and $c^i$ and $c^f$ are the initial and final entry pressures respectively).

Let $v_1$ and $v_2$ be fitting parameters for relative permeability changes. Then $k_{rw}$ and $k_{rn}$ are written as:

$$k_{rw} = \frac{\mathbb{F}(\bar{\chi}) S_{ew}^\Lambda}{1 - S_{ew} + \mathbb{F}(\bar{\chi}) S_{ew}^\Lambda}, \qquad k_{rn} = \frac{1 - S_{ew}}{1 - S_{ew} + \mathbb{F}(\bar{\chi}) S_{ew}^\Lambda}, \qquad (3)$$

where $\Lambda$ is a fitting parameter and $\mathbb{F}(\bar{\chi})$ is the dynamic function modeling the evolution from the initial to the final static curve ($\mathbb{F}(\bar{\chi}) = \min[\bar{\chi}(-v_1 C + v_2) + E^i, E^f]$, where $E^i$ and $E^f$ are the initial and final fitting parameters).

The mass conservation equations of component $k \in \{CO_2, \text{water}\}$ is written as:

$$\phi \sum_\alpha \frac{\partial(\rho_\alpha S_\alpha X_\alpha^k)}{\partial_t} + \sum_\alpha \nabla \cdot \left(\rho_\alpha X_\alpha^k \vec{u}_\alpha + \vec{j}_\alpha^k\right) = F^k, \qquad \alpha \in \{w, n\}, \qquad (4)$$

where $\phi$ is the porosity, $F^k$ the source term of component $k$, and $\rho_\alpha$ the density of phase $\alpha$. The multiphase extension of Darcy's law and Fick's law are written as:

$$\vec{u}_\alpha = -\frac{K k_{r\alpha}}{\mu_\alpha}(\nabla P_\alpha - \rho_\alpha \vec{g}), \qquad \vec{j}_\alpha^k = -\frac{(\phi S_\alpha)^{\frac{10}{3}} \rho_\alpha D_\alpha^k}{\phi^2} \nabla X_\alpha^k, \qquad (5)$$

where $K$ is the intrinsic permeability, $\vec{g}$ the gravitational vector, $D_\alpha^k$ the molecular diffusion coefficient, and $P_\alpha$ and $\mu_\alpha$ the pressure and viscosity of phase $\alpha$ respectively. To close the system of equations, we consider fugacity constraints $\left(f_w^k(P_w, T, X_w^k) - f_w^k(P_w, T, X_w^k) = 0\right)$ which model mass transfer between phases; and the fugacity coefficients are calculated as reported in Spycher and Pruess (2005).

In this work the system of interest is a two-dimensional caprock-aquifer flow system. The length of the reservoir is 4000 m, the height of the aquifer is 70 m, and height of the caprock is 30 m. The aquifer is located at a depth of 2600 m, initially saturated with brine ($S_w = 1$, $X_w^{CO_2}=5 \times 10^{-3}$), and in hydrostatic equilibrium. $CO_2$ is injected through the left side along the whole height of the aquifer at a rate of $10^{-3}$ kg/s for 10 years. The caprock and aquifer have contrasting properties. The permeability, porosity, and initial and final entry pressures in the caprock are $10^{-16}$ m², 0.1, and $10^6$ Pa and $10^4$ Pa while on the aquifer are $10^{-12}$ m², 0.2, and $10^4$ Pa and $10^2$ Pa, respectively. The WA dynamics in this system are assumed to be fast, i.e., the pore-scale parameter is set to $C = 10^{-5}$. The remaining model parameters have the following values (see Kassa et al. (submitted) for details): $b_1 = 10^{-7}$, $b_2 = 1.8$, $v_1 = 4.999 \times 10^{-3}$, $v_2 = 0.5$, $\lambda = 3.6$, $\Lambda = 1.3$, $E^i = 0.48$, $E^f = 3.37$, $S_{rw} = 0.2$, $D_\alpha^k = 2 \times 10^{-9}$ m²/s, $\rho_w = 1050$ kg/m³, $\rho_n = 716.7$ kg/m³, $\mu_w = 6.9 \times 10^{-4}$ Pa·s, $\mu_n = 5.9 \times 10^{-5}$ Pa·s, $T = 20°C$, $g = 9.81$ m/s², and $t_{ch} = 10^7$ s. The Open Porous Media (OPM) initiative allows to perform TPTC flow simulations through open-source code (Rasmussen et al., 2021). Link to complete codes for these numerical simulations can be found in https://github.com/daavid00/get2021.

## Results

To assess the WA effects, we compare the $CO_2$ front location (FL) and total $CO_2$ mass in the caprock neglecting and including WA effects. Figure 2 shows the simulated $CO_2$ saturation profile, FL, and total $CO_2$ mass in the caprock. In Figure 2a (neglecting WA effects), we observe that the $CO_2$ plume is only located in the aquifer and the FL almost reaches the right boundary after 10 years of injection in contrast to Figure 2b (including WA effects), where some $CO_2$ have migrated vertically into the caprock (Figure 2c) and the FL is closer to the injection well (Figure 2d).

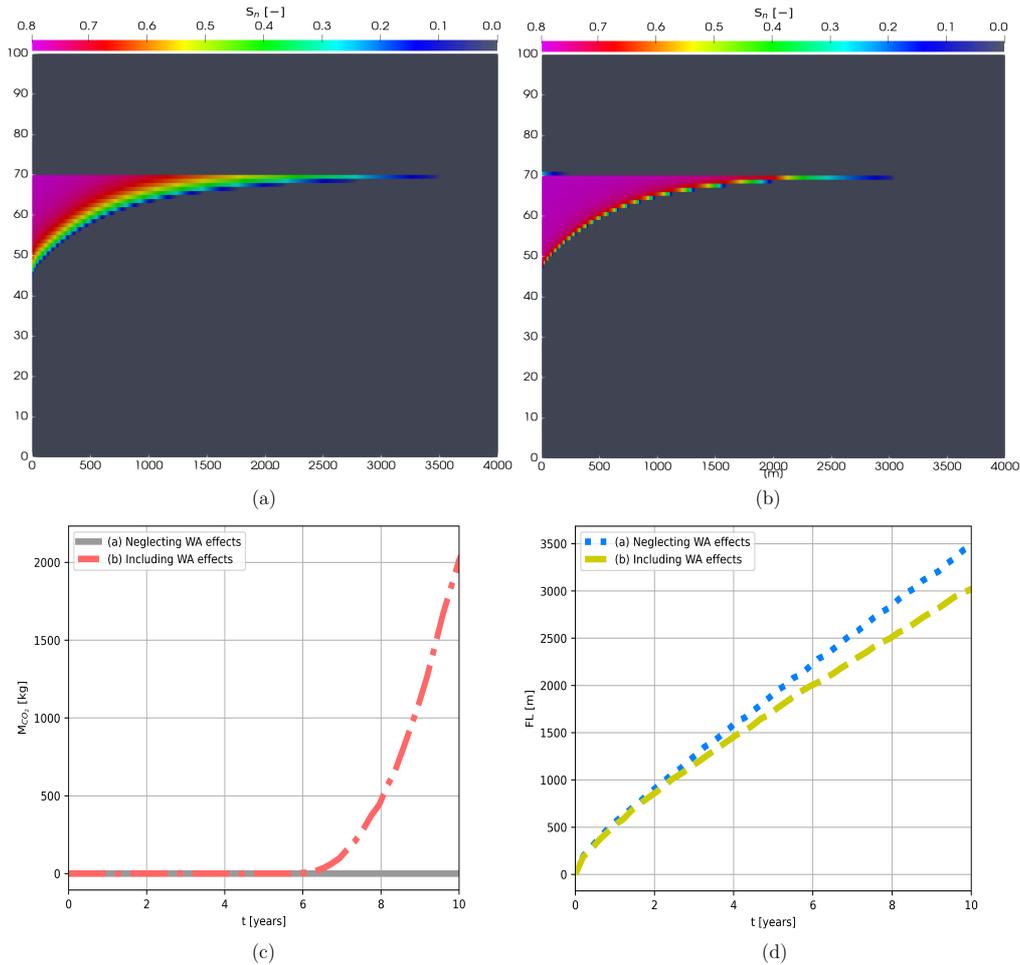

Figure 2: Spatial distribution of $CO_2$ (a) neglecting and (b) including WA effects; $CO_2$ (c) total mass in the caprock and (d) front location (FL) neglecting and including WA effects.

## Conclusions

Numerical studies are conducted in a two-dimensional aquifer-caprock system to assess long-term WA effects in $CO_2$ storage applications. The mathematical model is an extension of a two-phase two-component flow model by adding dynamic functions for capillary pressure and relative permeability relationships. This is achieved by including a new time-dependent variable, defined as upscaled exposure time, and a pore-scale rock parameter, into the aforementioned saturation functions. The implementation and numerical studies are performed in the OPM framework. Simulations on this 2D flow system confirm previous observations from simulations on isolated one-dimensional flow systems (i.e., 1D horizontal and vertical simulations, see Kassa et al. (submitted) for details). Particularly, long-term WA improves the aquifer storage efficiency.

**Acknowledgements**

This work was supported by the Research Council of Norway [grant number 255510] and CLIMIT-Demo/Gassnova [grant number 620073].